\documentclass[prb,twocolumn,amsmath,amssymb]{revtex4}

\usepackage{rotating} %for sidewaystable
\usepackage{graphicx} %figures

\newcommand{\dg}{\Delta G}
\newcommand{\dgb}{\Delta G_\text{bind}}
\newcommand{\dga}{\Delta G^\text{holo}_a}
\newcommand{\keq}{K_\text{eq}}
\newcommand{\xv}{\vec{x}}
\newcommand{\Xv}{\vec{X}}
\newcommand{\rv}{\vec{r}}
\newcommand{\refrv}{\vec{r}_*}
\newcommand{\refr}{r_*}
\newcommand{\Ua}{U_\text{a}}
\newcommand{\Ur}{U_\text{r}}
\newcommand{\pmf}{\mathcal{W}}

\begin{document}

\title{Absolute FKBP binding affinities obtained via non-equilibrium
unbinding simulations}
\author{F.\ Marty Ytreberg\footnote{E-mail: ytreberg@uidaho.edu}}
\affiliation{Department of Physics, University of Idaho, Moscow, ID 83844-0903}
\date{\today}

\begin{abstract}
    We compute absolute binding affinities for two ligands bound
    to the FKBP protein using non-equilibrium unbinding simulations.
    The methodology is straight-forward, requiring little or no modification
    to many modern molecular simulation packages.
    The approach makes use of a physical pathway, eliminating the
    need for complicated alchemical decoupling schemes.
    Results of this study are promising.
    For the ligands studied here the binding affinities are 
    typically estimated within less than 4.0 kJ/mol of the target values;
    and the target values are within less than 1.0 kJ/mol of experiment.
    These results suggest that non-equilibrium simulation could
    provide a simple and robust means to estimate
    protein-ligand binding affinities.
\end{abstract}

\maketitle

\section{Introduction}

The accurate estimation of binding affinities for protein-ligand systems
($\dg$) remains one of the most challenging tasks in
computational biophysics and biochemistry \cite{chipot-book}.
Due to the high computational cost of such free energy computation,
it is of interest to understand the advantages and limitations
of various $\dg$ methods.

Many previous studies \cite{kollman-affinity,hermans-restraint,gilson-binding1,gilson-binding2,wade-binding,vangunsteren-estrogen,chipot-binding,roux-gsbp,karplus-binding,vangunsteren-bindingref,mccammon-fkbp,shirts-fkbp,pearlman-pbsa,aqvist-binding,roux-wham,roux-decoupling,roux-fkbp,mobley-symmetry,shirts-fkbp2,olson-2006,mobley-restraint,olson-2008}
have calculated protein-ligand binding affinities using
equilibrium free energy methods such as
thermodynamic integration \cite{kirkwood},
free energy perturbation \cite{zwanzig,valleau},
and weighted histogram analysis \cite{swendsen-wham}.
Due to the introduction of the novel Jarzynski approach
\cite{jarzynski} it is also possible
to estimate $\dg$ from non-equilibrium simulations.
However, the estimation of $\dg$ for protein-ligand
binding using non-equilibrium approaches
remains largely untested.
Two recent studies used non-equilibrium simulations to
test unbinding pathways \cite{mccammon-jarz,abrams-pmf}.
Studies by other groups found that estimating
$\dg$ via non-equilibrium simulations resulted in a large error compared
to experiment \cite{grubmuller-jarz,michielin-pmf}.
A recent study by Kuyucak and collaborators demonstrated
that use of non-equilibrium simulation required longer simulation times
than umbrella sampling \cite{kuyucak-pmf}.

In this report, apparently for the first time,
we demonstrate the ability to compute accurate
(as compared to experimental data)
protein-ligand binding $\dg$ estimates following
a non-equilibrium methodology.
The approach relies on performing multiple
non-equilibrium unbinding simulations using a physical pathway,
i.e., pulling the ligand out of the binding pocket,
and then uses the Jarzynski relation \cite{jarzynski}
to estimate $\dg$.
The system is an FKBP protein complexed with 4-hydroxy-2-butanone (BUQ)
and dimethyl sulfoxide (DMSO).
The motivation for using this system is that comparison
to experiment is possible \cite{walkinshaw-fkbp}
and many previous computational studies have been performed
\cite{mccammon-fkbp,shirts-fkbp,shirts-fkbp2,roux-fkbp,olson-2006}.

The importance of pursuing non-equilibrium
methods such as used in this report is three-fold:
(i) The approach is trivially parallelizeable
since each non-equilibrium unbinding simulation is performed independently.
(ii) The method is simple to implement in many
existing simulation packages such as GROMACS \cite{gromacs}, used here;
little or no modification to the code is necessary.
(iii) Since a physical pathway is utilized,
there is no need to use alchemical decoupling schemes.

This study represents the first stage of a project comparing
the efficiencies of various free energy methods for
protein-ligand $\dg$ computation.
We note that efficiency studies have been carried out for
other non-protein systems \cite{kollman-efficiency,kofke-efficiency,hummer,shirts-efficiency,vangunsteren-efficiency,ytreberg-efficiency}.

\section{Theory}

In general, the absolute binding affinity is defined as the
free energy difference between the unbound (apo) and bound (holo) states
of the protein-ligand system.
We define the apo state as when the protein and ligand
are not interacting due to a large separation between them.
The holo state is defined to be when the ligand is in the binding
pocket of the protein.
Experimentally, the binding affinity is measured by determining
the equilibrium constant $\keq=[PL]/[P][L]$, where $[PL]$ denotes
the concentration of the protein-ligand complex,
and $[P]$ and $[L]$ are the concentrations of the
apo protein and free ligand, respectively.
Then the absolute binding affinity is given by
$\dgb=-k_BT\ln(\keq/V_0)$,
where $k_B$ is the Boltzmann constant,
$T$ is the system temperature, and $V_0$ is the standard volume
used for the experiment (typically $V_0=1.661$ nm$^2$
corresponding to 1.0 mol/liter concentration).

\begin{figure}
    \begin{center}
	\includegraphics[width=2.0in]{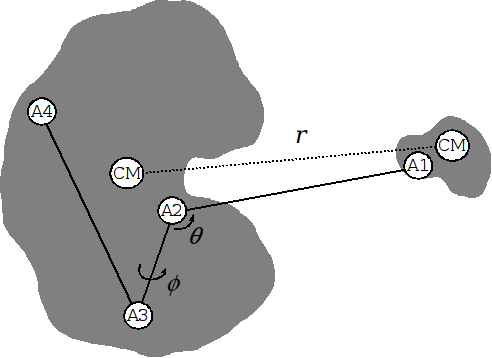}
    \end{center}
    \caption{\label{fig-coords}
    The coordinate system used for the restraints $\Ua(\theta,\phi)$
    and $\Ur(r,t)$.
    The value of $r$ is given by the center of mass separation between
    the protein and ligand.
    A1-A4 are the heavy atoms used to define the coordinate system for $\Ua$.
    For the FKBP-DMSO system A1 = DMSO:S1, A2 = TRP59:N, A3 = HIS25:N
    and A4 = ALA64:N.
    For the FKBP-BUQ system A1 = BUQ:C2, and A2-A4 are the same as FKBP-DMSO.
    } 
\end{figure}

Following the notation of Roux and collaborators \cite{roux-wham,roux-fkbp}
(also see discussion in
Refs.\ \onlinecite{chipot-book,gilson-binding1,karplus-binding})
the equilibrium constant is given by a ratio of integrals over
the apo and holo regions of configurational space
\begin{equation}
    \keq=\frac
    {\int_\text{holo}d\xv\int d\Xv\;e^{-\beta U(\xv,\Xv)}}
    {\int_\text{apo}d\xv\;\delta(\rv-\refrv)\int d\Xv\;e^{-\beta U(\xv,\Xv)}}\,,
    \label{eq-keq}
\end{equation}
where $\beta=1/k_BT$, $\xv$ represents the configurational
coordinates of the ligand, $\Xv$ are the coordinates
of the protein and solvent,
and $U(\xv,\Xv)$ is the potential energy of the system.
The vector $\rv$ defines the location of the center of mass of the ligand
relative to the center of mass of the protein (see Fig.\ \ref{fig-coords}),
and $\refrv$ is a reference value taken to be when the ligand
and protein are not interacting.

Equation \eqref{eq-keq} can be used to compute binding affinities
using various computational strategies.
In our study, we restrain the ligand relative to the protein
so that the ligand remains along the binding axis.
The potential energy of this axial restraint is given by
$\Ua(\theta,\phi)=\frac{1}{2}k_a\big[(\theta-\theta_0)^2+(\phi-\phi_0)^2\big]$,
where $k_a$ is the force constant
and $\theta_0,\phi_0$ are reference values of the coordinates;
see Fig.\ \ref{fig-coords}.
With this restraint defined, Eq.\ \eqref{eq-keq}
can be written as a product of dimensionless ratios of integrals
\begin{align}
    e^{-\beta \dgb}&=\frac{\keq}{V_0} \equiv I_1 \times I_2\,, \nonumber \\
    I_1&=\frac{\int_\text{holo}d\xv\int d\Xv\;e^{-\beta U}}
	{\int_\text{holo}d\xv\int d\Xv\;e^{-\beta(U+\Ua)}}\,,
    	\nonumber \\
    I_2&=\frac{1}{V_0}\;\frac{\int_\text{holo}d\xv
	    \int d\Xv\;e^{-\beta(U+\Ua)}}
	{\int_\text{apo}d\xv\;\delta(\rv-\refrv)
	    \int d\Xv\;e^{-\beta U}}\,,
    \label{eq-keq2}
\end{align}
where strategies for computing
the terms $I_1$ and $I_2$ will be discussed below.

The first term $I_1\equiv e^{-\beta\dga}$ in
Eq.\ \eqref{eq-keq2} corresponds to the free energy difference
associated with restraining the protein to the binding
axis while in the holo state.
This free energy can be computed using any standard technique
by performing simulations for a range of force constants
from 0 to $k_a$.
For the current study we chose to compute this free energy difference
via a multi-stage Bennett approach \cite{bennett}.

To determine the second term $I_2$ in Eq.\ \eqref{eq-keq2}
we define the potential of mean force (PMF) $\pmf$,
with the restraint potential $\Ua$ present,
as a function of the \emph{scalar} distance $r$
\begin{equation}
    e^{-\beta[\pmf(r_2)-\pmf(r_1)]}=
    \frac{\int d\xv\;\delta(r-r_2) \int d\Xv\;e^{-\beta(U+\Ua)}}
    {\int d\xv\;\delta(r-r_1) \int d\Xv\;e^{-\beta(U+\Ua)}}\,.
\end{equation}
Integrating the PMF over both apo and holo
regions we can obtain
\begin{align}
    \int_\text{apo} d\rv_1 \;\delta(\rv_1-\refrv)\;e^{-\beta\Ua}
	\int_\text{holo} dr_2 \;e^{-\beta[\pmf(r_2)-\pmf(r_1)]} \nonumber\\
    =\frac{1}{4\pi\refr^2}
	\frac{\int_\text{holo}d\xv\int d\Xv\;e^{-\beta (U+\Ua)}}
	{\int_\text{apo}d\xv\;\delta(\rv-\refrv)\int d\Xv\;e^{-\beta U}}\,,
    \label{eq-I2}
\end{align}
where we have used the fact that 
$\delta(r-\refr)=4\pi r^2\,\delta(\rv-\refrv)$ for the apo integral
since the PMF is independent of the direction of $\rv$ when the ligand
is not interacting with the protein.
Thus $I_2$ can be evaluated by estimating the integral of the PMF
in Eq.\ \eqref{eq-I2},
\begin{widetext}
\begin{align}
    I_2&=\frac{4\pi\refr^2}{V_0}
	\int_\text{apo} d\rv_1 \;\delta(\rv_1-\refrv)\;e^{-\beta\Ua}
	\int_\text{holo} dr_2 \;e^{-\beta[\pmf(r_2)-\pmf(r_1)]} \nonumber\\
    &=\frac{4\pi\refr^2}{V_0}
	\int_\text{apo} r_1^2 dr_1 \cos\theta_1 d\theta_1 d\phi_1\;
	\frac{\delta(r_1-\refr)}{4\pi r_1^2}
	\;e^{-\beta\Ua(\theta_1,\phi_1)}\;e^{+\beta\pmf(r_1)}
	\int_\text{holo} dr_2\;e^{-\beta\pmf(r_2)} \nonumber\\
    &=\frac{\refr^2}{V_0}\;e^{+\beta\pmf(\refr)}
	\int_\text{apo} \cos\theta_1 d\theta_1 d\phi_1
	\;e^{-\beta\Ua(\theta_1,\phi_1)}
	\int_\text{holo} dr_2 \;e^{-\beta\pmf(r_2)}.
\end{align}
\end{widetext}
Below the apo integral will be evaluated analytically
and the holo integral will be evaluated using quadrature.

With our approximations above the absolute binding free
energy can now be estimated via the relation
\begin{align}
    \dgb&=\dga-\pmf(\refr)-k_BT \ln\Bigg[\frac{\refr^2}{V_0}\times \nonumber \\
	&\int_\text{apo} \cos\theta d\theta d\phi
	\;e^{-\beta\Ua(\theta,\phi)}
	\int_\text{holo} dr \;e^{-\beta\pmf(r)}\Bigg].
    \label{eq-dgbind}
\end{align}
This is our central theoretical result.
Thus, estimating $\dgb$ in the current framework requires
three computations: the PMF must be calculated (detailed below),
$\dga$ must be approximated, and the apo integral must be
analytically evaluated.

One may note that a different, yet viable, non-equilibrium approach would be
to set $k_a=0$, i.e., remove the axial restraint. 
This would simplify the $\dgb$ calculation since $\dga=0$ and the
apo integral would be $4\pi$ in Eq.\ \eqref{eq-dgbind},
and thus only the PMF would be needed.
Further, the binding axis would not need to be defined by the researcher,
which would allow the ligand more flexible unbinding routes.
However, there are two important advantages to using the axial restraint.
Most important is that since the axial restraint limits the 
allowable configurational space for the ligand,
the PMF converges much more quickly than with no restraint.
Also, in cases where the binding pocket is not at the protein surface,
or when more than one viable pathway exists,
it may be advantageous, or even necessary, to define the binding
path to obtain meaningful results.

\subsection{Estimating the PMF}

The PMF will be computed using two different approaches:
the Hummer-Szabo method, and the stiff-spring approximation
with the second cumulant expansion.
Below we summarize these techniques.

In the Hummer-Szabo approach the PMF is estimated
by performing multiple non-equilibrium pulling simulations
along the reaction coordinate $r$
by using a time-dependent biasing potential $\Ur(r,t)=k_r[r-(r_0+vt)]^2$,
where $k_r$ is the force constant, $r=r(t)$ is the
protein-ligand center of mass separation,
$r_0$ is an initial reference separation
which is constant for all pulling simulations (i.e., $r_0\neq r(0)$),
and $v$ is the speed at which the biasing center is moved.
The PMF is then estimated via \cite{hummer-szabo-2001,hummer-szabo-2005}
\begin{equation}
    e^{-\beta\pmf(r)}=
    \dfrac
    {\sum_t\dfrac
	{\langle\delta[r-r(t)]\;\exp(-\beta W_t)\rangle}
	{\langle\exp(-\beta W_t)\rangle}}
    {\sum_t\dfrac{\exp(-\beta \Ur(r,t))}{\langle\exp(-\beta W_t)\rangle}}
    \times e^{-2\ln(r)}\,,
    \label{eq-hz}
\end{equation}
where the sum is over time slices $t$,
and the $2\ln(r)$ term is the Jacobian correction which is necessary
since $r$ is a radial distance \cite{vangunsteren-pmf}.
The $\langle\ldots\rangle$ indicates an ensemble average for
pulling simulations drawn from the Boltzmann distribution
corresponding to the initial system potential energy
$U_\text{tot}(\xv,\Xv,t=0)=U(\xv,\Xv)+\Ua(\theta,\phi)+\Ur(r,0)$.
The work for a given time slice is given by \cite{hummer-szabo-2001}
\begin{align}
    W_t&=-\int \frac{\partial U_r(r,t)}{\partial r}dr-\Ur(r(0),0)
    \nonumber\\
    &=k_rv\bigg[\frac{1}{2}vt^2-\int_0^t [r(t')-r_0]dt'\bigg] - \Ur(r(0),0)\,.
    \label{eq-whz}
\end{align}
Note that $W_t$ is the accumulated work minus the initial $t=0$
biasing energy.

The stiff-spring approximation utilizes the well-known Jarzynski equality
\cite{jarzynski,jarzynski-pre,crooks-pre} to estimate the PMF.
The approximation is that for a sufficiently large force constant $k_r$
that the protein-ligand separation closely follows the biasing center, i.e.,
$\xi \equiv r_0+vt \approx r$.
Park and Schulten thus concluded that
the accumulated work along the reaction coordinate $r$
is approximately equal to the accumulated work for a given time slice
\cite{schulten-jarz,schulten-jarz2}
\begin{equation}
    e^{-\beta\pmf(r)}\approx
    e^{-\beta\dg(r)}=\big\langle e^{-\beta W_r} \big\rangle
    \times e^{-2\ln(r)}\,,
    \label{eq-ss}
\end{equation}
where the $2\ln(r)$ term is necessary due to the Jacobian correction
\cite{vangunsteren-pmf}
and the work is determined by integrating the biasing force
over the location of the bias center 
\begin{align}
    W_r \approx W_\xi &= -\int_{r_0}^{r_0+vt}
	\frac{\partial U_r(r,t)}{\partial r}\;
	\frac{\partial r}{\partial\xi}\;d\xi-\Ur(r(0),0) \nonumber\\
    &\approx -\int_{r_0}^{r_0+vt}
	\frac{\partial U_r(r,t)}{\partial r}\;d\xi-\Ur(r(0),0)\,,
\end{align}
where we have used the fact that $\partial r/\partial \xi\approx 1$.
Applying the cumulant expansion to Eq.\ \eqref{eq-ss},
we obtain the final expression used estimate the PMF for the stiff-spring
approach \cite{schulten-jarz,schulten-jarz2}
\begin{equation}
    \pmf(r) \approx \langle W_r \rangle
	-\frac{\beta}{2}[\langle W_r^2\rangle-\langle W_r \rangle^2]
	+2\ln(r)\,.
    \label{eq-gauss}
\end{equation}

For the results given in this report the ligand
is pulled out of the binding pocket, 
and the reverse process of pulling the ligand into the pocket
is not considered.
Future studies will include the reverse process since
the use of bi-directional simulation
has been shown to be an effective approach to accurate $\dg$ estimation
\cite{bennett,lu-2003,shirts-prl,lu-jcc,lu-wu,shirts-efficiency,ytreberg-efficiency,minh-prl-2008}.

We note two aspects of the relationships embodied in
Eqs.\ \eqref{eq-hz} and \eqref{eq-gauss}:
(i) The equality in Eq.\ \eqref{eq-hz} holds only in the case of obtaining
all possible pulling trajectories.
The approximation in Eq.\ \eqref{eq-gauss} is an equality for the case
that the work value distribution is perfectly Gaussian.
Thus, it is important to calculate uncertainty
estimates for the PMF, and if possible, to compare results to
an independent computational measure---below we will compare our results
to use of umbrella sampling.
(ii) The relation is independent of the speed at which the system is
forced, i.e., the unbinding speed.
In practice, however, it has been found that
the speed chosen can dramatically affect the convergence behavior
of the estimates
\cite{hummer,bustamante-bias,ytreberg-efficiency}.

\subsection{Use of a physical pathway}

It is useful to consider the advantages and disadvantages of
using a physical (rather than alchemical) pathway.
The regions of configurational space corresponding to apo and holo
in Eq.\ \eqref{eq-keq} are well-separated with no overlap, thus
a pathway connecting them is typically created.
For our discussion below, this pathway will be parameterized using
the variable $\lambda$.

In the case of a physical pathway, such as in the current study,
$\lambda=r$ represents the protein-ligand separation.
By contrast, for an alchemical pathway $\lambda$ is generally
a parameter that scales the strength of the interactions between the ligand
and rest of the system.

Our use of a physical pathway is motivated by several factors.
Alchemical pathways are typically much more difficult to implement
than physical pathways since interactions must be scaled carefully.
In addition, restraints must often be employed such that the
non-interacting parts do not drift away from the region of interest.

We note that there are disadvantages to using physical pathways.
Physical pathways may require the researcher to determine the pulling
direction such that the ligand exits the binding pocket, i.e., determined
by choice of $\Ua$ in this report.
Alchemical pathways do not require such a choice.
Perhaps most important, physical pathways require
larger system sizes when explicit solvent is used, as in the current report.
The size of the system must be large enough that the ligand can be pulled
to a distance such that interactions between 
the ligand and protein are negligible.

In cases where the binding site is buried deep within the protein,
alchemical methods should be much more efficient than physical
approaches.
However, when the binding pocket is close to the protein surface,
as for the current study, it is not clear where alchemical or physical
approaches are more efficient and/or accurate.

Another important consideration is that the use of a physical pathway
allows the researcher to determine the PMF along the pathway.
This PMF can give insights into binding that are simply not possible
when using alchemical methods, e.g., determining the preferred
binding pathway when multiple pathways are present
\cite{mccammon-jarz,abrams-pmf}.

\subsection{Use of a non-equilibrium approach}

Non-equilibrium approaches, such as used in the current study,
rely on computing the work required to force the system from
one state to the other rapidly enough that equilibrium is not
attained at any value of $\lambda$.
This process is repeated many times and the resulting
distribution of work values is used to estimate $\dg$ \cite{jarzynski}.
By contrast, equilibrium free energy methodologies such as
thermodynamic integration \cite{kirkwood},
free energy perturbation \cite{zwanzig,valleau},
and weighted histogram analysis \cite{swendsen-wham},
share the common strategy of
generating equilibrium ensembles of configurations for multiple
values of the scaling parameter $\lambda$.
It is important when performing such $\dg$ computation that
enough simulation time is spent to equilibrate at each value of $\lambda$
such that the resulting ensemble is valid for the current $\lambda$.

It is not currently known whether equilibrium or non-equilibrium
methodologies offer an efficiency advantage for typical protein-ligand
binding affinity computation.
Equilibrium methods have been widely used to generate
accurate $\dg$ estimates for protein-ligand binding
\cite{kollman-affinity,hermans-restraint,gilson-binding1,gilson-binding2,wade-binding,vangunsteren-estrogen,chipot-binding,roux-gsbp,karplus-binding,vangunsteren-bindingref,mccammon-fkbp,shirts-fkbp,pearlman-pbsa,aqvist-binding,roux-wham,roux-decoupling,roux-fkbp,mobley-symmetry,shirts-fkbp2,olson-2006,mobley-restraint,olson-2008}.
However, if equilibrium is not attained
the resulting $\dg$ estimate can be heavily biased.
With few very recent exceptions
\cite{mccammon-jarz,grubmuller-jarz,abrams-pmf,michielin-pmf,kuyucak-pmf}
non-equilibrium methods are largely untested on protein-ligand systems.
In previous calculations of relative solvation free energies
non-equilibrium methods were proven to be equal or superior
in efficiency to commonly used equilibrium methods \cite{ytreberg-efficiency}.

A key advantage of non-equilibrium methodologies is the ease
that one can parallelize the $\dg$ calculation.
Since each work value must necessarily be generated independently,
the corresponding simulations can be run in parallel with
no loss of accuracy to the final $\dg$ estimate.
Equilibrium $\dg$ computations, by contrast, are not trivially parallelizeable.
One can imagine performing each $\lambda$ simulation in parallel,
however one must be very careful about the configurations used
to start each $\lambda$ simulation.
In typical cases it is necessary to start the current $\lambda$ simulation
using the final snapshot from the previous $\lambda$ simulation; thus,
the $\lambda$ simulations are performed in a serial fashion.
If this is not done, the amount of time needed to equilibrate at
each value of $\lambda$ could be heavily dependent on the
chosen starting structures.
The $\dg$ estimate could be heavily biased if the time spent
for equilibration at each $\lambda$ value is inadequate.

\section{Methods}

\subsection{Computational details}

The initial coordinates for the FKBP-ligand complexes
were obtained from the Protein Data Bank \cite{pdb}:
1D7H for FKBP-DMSO, and 1D7J for FKBP-BUQ.
The topologies for DMSO and BUQ were then generated by the
PRODRG server \cite{prodrg}, with partial charges
modified by the author.

The GROMACS simulation package version 3.3.3 \cite{gromacs}
was used with the default GROMOS-96 43A1 forcefield \cite{gromos}.
The software was slightly modified to provide
the biasing potential $\Ur$ which depends only on the center of
mass separation between the ligand and the protein.
Protonation states for the histidine residues were selected
by the GROMACS program {\tt pdb2gmx}:
HIS25 was protonated at N$\delta$1,
and HIS87 and HIS94 were protonated at N$\epsilon$2.
The protein-ligand complexes were then solvated in a cubic box of
SPC water \cite{spc} of approximate initial size 6.8 nm a side.
A single chloride ion was randomly placed in each water box
to give a net neutral charge, and then each system was
minimized using steepest decent for 500 steps.
To allow for equilibration of the water, each system was then simulated
for 1.0 ns with the positions of all heavy atoms in the ligand and protein
harmonically restrained with a force constant of 1000 kJ/mol/$\text{nm}^2$.
The temperature was maintained at 300 K using
Langevin dynamics \cite{langevin} with a friction coefficient of 1.0 amu/ps.
The pressure was maintained at 1.0 atm using the
Berendsen algorithm \cite{berendsen}.
We note that the Berendsen algorithm does not produce 
canonically distributed structures, however, none
of the resulting simulation frames were used for generating
$\dg$ estimates, as will be seen below.
The LINCS algorithm \cite{lincs}
was used to constrain hydrogens to their ideal lengths
and heavy hydrogens were used---the hydrogen mass was increased by a factor
of four and this increase was subtracted from the bonded heavy atom
so that the mass of the system remained unchanged---allowing the
use of a 4.0 fs timestep.
Particle mesh Ewald \cite{pme}
was used for electrostatics with a real-space cutoff of 1.0 nm
and a Fourier spacing of 0.1 nm.  Van der Waals interactions used a
cutoff with a smoothing function such that the interactions smoothly
decayed to zero between 0.75 nm and 0.9 nm.
Dispersion corrections for the energy and pressure were utilized
\cite{allen-tildesley}.

After the position restrained simulation,
a 4.0 ns equilibrium simulation at constant temperature
and volume, with restraints,
was used to generate starting configurations for use in
the PMF calculations.
Each FKBP-protein complex was simulated
with parameters chosen identical to the position restrained simulation
above except that the volume was fixed at the value of the final
configuration from the position restrained simulations.
Importantly, for Eq.\ \eqref{eq-hz} and \eqref{eq-gauss}
to be used these equilibrium simulations must include
the restraints, i.e., both $\Ua$ and $\Ur$ were present.
For both the DMSO and BUQ systems the axial restraint
used a force constant $k_a=1000$ kJ/mol/$\text{nm}^2$,
and $\theta_0,\phi_0$ were chosen to be equal to the values
from the final snapshot of the position restrained simulation.
For both systems the biasing potential $\Ur$ used a force constant
$k_r=3000$ kJ/mol/$\text{nm}^2$ and $r_0=0.5$ nm
and a speed $v=0$.

Starting structures for the unbinding simulations were chosen
to be equally spaced within the 4.0 ns equilibrium simulation.
So, if 40 starting structures were desired, then the spacing
between snapshots was 100 ps.
The pulling simulations were performed using identical parameters
to the 4.0 ns equilibrium simulation,
except that the bias center was moved at a constant speed $v$
ranging from $1.0\times 10^{-4}$ nm/ps to $8.0\times 10^{-4}$ nm/ps.
The pulling simulations were discontinued when the bias center
was at a position of 2.5 nm.

\begin{figure}
    \begin{center}
	\includegraphics[width=3in]{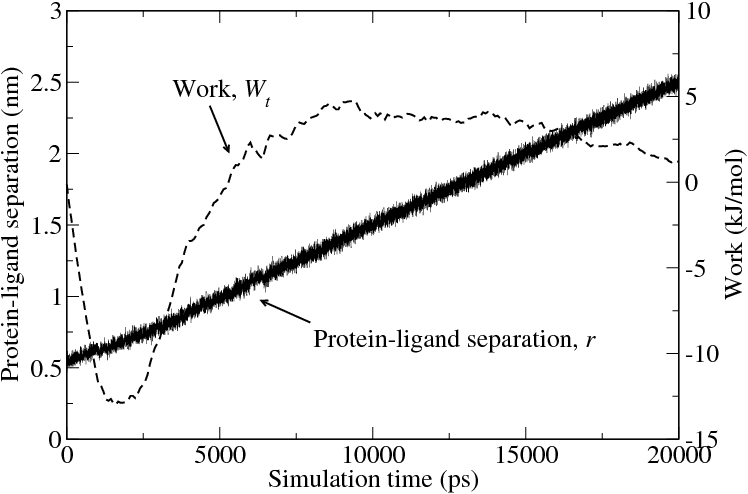}
    \end{center}
    \caption{\label{fig-work}
    Results shown here are for a single non-equilibrium pulling simulation
    performed on the FKBP-DMSO system
    using a pulling speed of $1.0\times10^{-4}$ nm/ps.
    The solid line shows the protein-ligand center of mass separation
    as a function of simulation time.
    The dashed line shows the
    accumulated work $W_t$ as computed by Eq.\ \eqref{eq-whz}
    as a function of simulation time.
    }
\end{figure}

The non-equilibrium unbinding simulations provided us with
the protein ligand separation $r$ at every time step,
which we used to compute the work and thus the resulting PMF.
An example of this is shown in Fig.\ \ref{fig-work} where
we have used Eq.\ \eqref{eq-whz}
to compute the work as a function of simulation time $W_t$.

\subsection{Computing $\dga$}

We used the Bennett acceptance ratio approach
to compute the free energy differences $\dga$ associated with
the axial restraints \cite{bennett}.
With the ligand bound to the protein we performed 1.0 ns equilibrium
simulations for each of the values
$k_a=0,25,40,60,90,150,200,300,450,700,1000$.
The first 0.5 ns of each simulation were discarded for equilibration,
and the remaining 0.5 ns were used to compute $\dga$.
We did not attempt to optimize efficiency of the
$\dga$ computations, our only concern was accurate values,
so it may be possible to reduce the
total computational time from that described above.

\subsection{Uncertainty estimation}

We estimated the uncertainty in our $\dgb$ estimates using
the bootstrap approach applied to the PMF:
(i) The reference value of the PMF given by $\pmf(\refr)$
was computed via Eq.\ \eqref{eq-gauss}
using $N$ work values chosen at random (with replacement) from a dataset
containing $N$ values;
(ii) The above step was repeated until the mean and standard deviation
of the free energy estimates converged; around
100,000 trials in our study.
(iii) The uncertainty is given by the converged standard deviation of the
free energy estimates.

For comparison, we also used the uncertainty analysis obtained by
Zuckerman and Woolf \cite{zuckerman-bias}, and the Bustamante group
\cite{bustamante-bias}.
These uncertainty estimates are reported to be
accurate when the variance in the
estimate dominates over the bias (as in the case of large $N$).

\subsection{Generating a target PMF}

Since the purpose of the current study was to test the effectiveness of
non-equilibrium strategies it is important to have an independent estimate
of the PMF.
Thus, we computed the PMF using umbrella sampling and
weighted histogram analysis (WHAM) \cite{swendsen-wham}.
Simulations were performed with the restraints $\Ua$ and $\Ur$.
For the umbrella sampling simulations the speed was set to $v=0$,
all other parameters were identical to the non-equilibrium simulations,
and 41 windows were used $r_0=0.50,0.55,0.60,...,2.45,2.50$.
Each window was simulated for a total time of 12 ns;
6 ns were discarded for equilibration and 6 ns were used for
the WHAM analysis.
Thus, the total simulation time was nearly three times greater than
the non-equilibrium simulations detailed above.
No attempt was made to test the efficiency since the goal was to
generate the most accurate PMF.
Note that the $2\ln(r)$ Jacobian correction from Eqs.\ 
\eqref{eq-hz} and \eqref{eq-gauss} were also used for the target PMF.

\section{Results and Discussion}

The results of this study are very encouraging.
Using the non-equilibrium methodology outlined above we estimated
the the binding affinity for the FKBP-DMSO and FKBP-BUQ complexes
typically within less than 4.0 kJ/mol of the target values; and the target
values are within less than 1.0 kJ/mol of experiment.

\begin{figure}
    \begin{center}
	\includegraphics[width=1.6in]{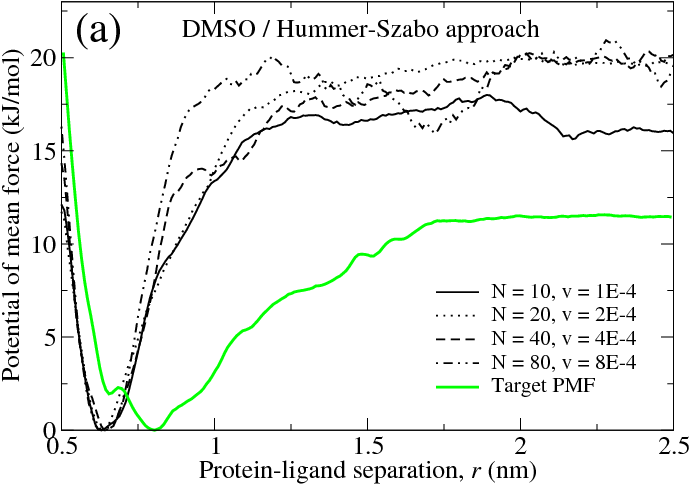}
	\hspace{2pt}
	\includegraphics[width=1.6in]{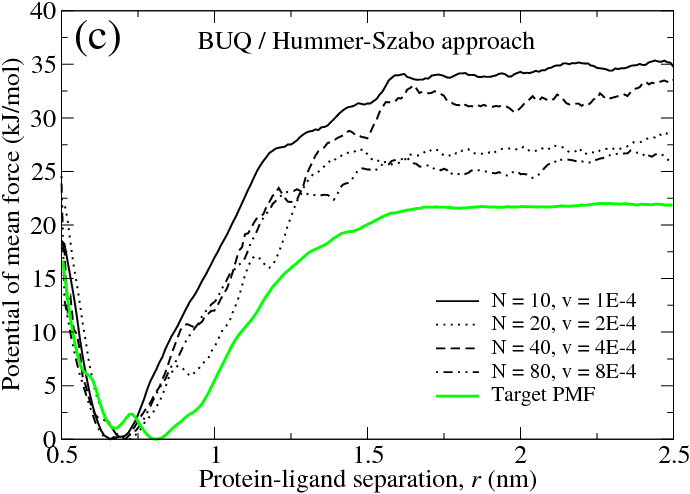}\\
	\vspace{2pt}
	\includegraphics[width=1.6in]{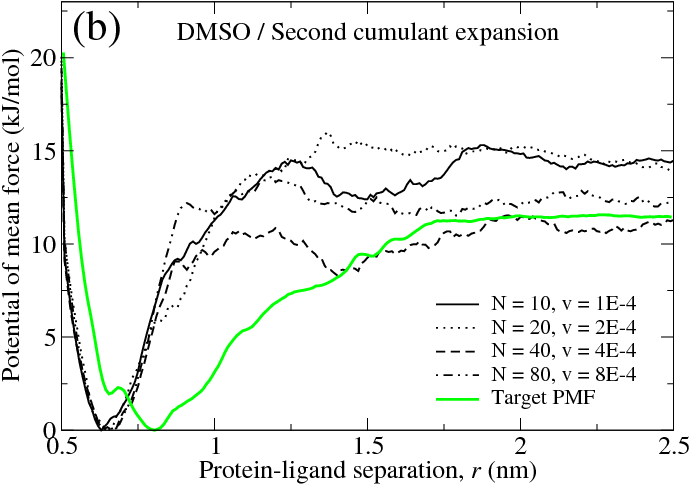}
	\hspace{2pt}
	\includegraphics[width=1.6in]{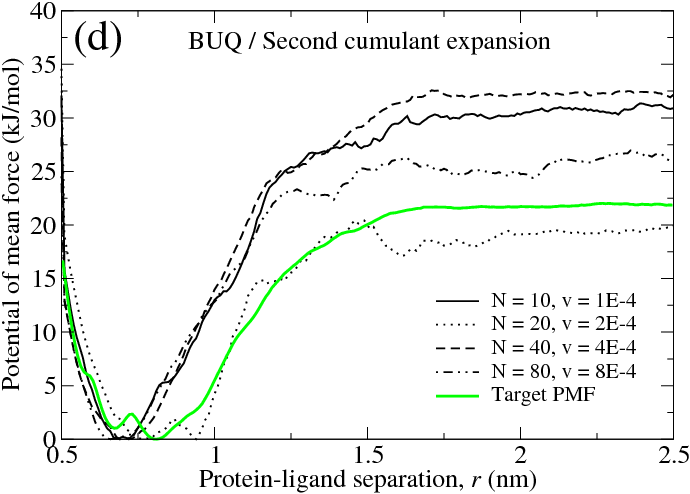}
    \end{center}
    \caption{\label{fig-pmf}
    (color online)
    The PMF as a function of the protein-ligand center of mass separation.
    These PMF curves were numerically integrated for
    Eq.\ \eqref{eq-dgbind} and used to generated the $\dgb$ estimates
    shown in Table \ref{tab-fkbp}.
    All non-equilibrium results utilized the same
    total amount of simulation time.
    For all plots the light colored solid curve shows the target PMF
    generated via equilibrium umbrella sampling and WHAM.
    The target PMF curves utilized three times more simulation time
    as the non-equilibrium simulations,
    and thus we are not attempting to compare the accuracy of equilibrium
    and non-equilibrium approaches.
    (a) FKBP-DMSO system using the Hummer-Szabo approach of
    Eq.\ \eqref{eq-hz}.
    (b) FKBP-DMSO system using the stiff-spring second cumulant
    expansion approximation of Eq.\ \eqref{eq-gauss}.
    (c) FKBP-BUQ system using the Hummer-Szabo approach of
    Eq.\ \eqref{eq-hz}.
    (d) FKBP-BUQ system using the stiff-spring second cumulant
    expansion approximation of Eq.\ \eqref{eq-gauss}.
    }
\end{figure}

Figure \ref{fig-pmf} shows the PMF as a
function of protein-ligand separation for all systems studied here.
Data is shown for both DMSO and BUQ systems, with
pulling speeds indicated on each plot.
Note that the same amount of total simulation time was spent on each
non-equilibrium PMF estimate, but that the target PMF utilized three
times as much simulation as the non-equilibrium estimates,
thus we are not attempting to compare non-equilibrium and equilibrium
approaches in this study.
For both systems the non-equilibrium estimates tend to overestimate
the target PMF, and underestimate the broadness of the PMF minimum.
This suggests that the pulling speeds were not slow enough to properly
sample the PMF.
Also, use of stiff-spring approximation with the second cumulant
expansion does tend to improve the non-equilibrium PMF curves.

\begin{table*}
    \begin{center}
    \begin{tabular}{lccccccccc}
    \hline\hline
    Ligand & $N$ & Speed (nm/ps) & $\sigma_W$ & $\dgb$$^a$ & $\dgb$$^b$
	& Uncty$^c$ & Uncty$^d$	& Target$^e$ & Exp \\
    \hline
    DMSO & 10 & $1.0\times 10^{-4}$ & 6.9  & -12.6 & -11.1 & 3.5 & 2.4
	    & -9.2 & -9.7 \\
         & 20 & $2.0\times 10^{-4}$ & 7.9  & -15.6  & -10.7 & 1.2 & 1.0 & & \\
	 & 40 & $4.0\times 10^{-4}$ & 9.4  & -16.3 & -8.2 & 1.6 & 1.5 & & \\
         & 80 & $8.0\times 10^{-4}$ & 10.4 & -15.9 & -9.2 & 2.2 & 1.7 & & \\
    \hline
    BUQ & 10 & $1.0\times 10^{-4}$ & 7.8  & -30.0 & -26.7 & 3.1 & 2.2
	    & -18.3 & -18.9 \\
        & 20 & $2.0\times 10^{-4}$ & 11.9 & -23.2 & -16.0 & 4.7 & 2.4 & & \\
	& 40 & $4.0\times 10^{-4}$ & 8.1  & -28.4 & -28.2 & 3.8 & 2.4 & & \\
        & 80 & $8.0\times 10^{-4}$ & 11.3 & -36.0 & -22.3 & 1.7 & 1.5 & & \\
    \hline\hline
    \end{tabular}
    \end{center}
    \caption{\label{tab-fkbp}
    Comparison between non-equilibrium binding affinity estimates and
    target and experimental binding affinities.
    All energy values are shown in units of kJ/mol.
    The first column describes the ligand used.
    The second column contains the number of work values
    $N$ used in the estimate, and the third and fourth columns are
    respectively,
    the corresponding speed of the restraint attached to the ligand,
    and the standard deviation of the work values.
    The rightmost column gives the experimental results reported in
    Ref.\ \onlinecite{walkinshaw-fkbp}.\\
    $^a$ Binding affinity estimate obtained via the Hummer-Szabo approach
    using Eqs.\ \eqref{eq-dgbind} and \eqref{eq-hz}.\\
    $^b$ Binding affinity estimate obtained using the stiff-spring
    second cumulant expansion approximation using
    Eqs.\ \eqref{eq-dgbind} and \eqref{eq-gauss}.\\
    $^c$ Uncertainty estimate computed via the bootstrap method.\\
    $^d$ Uncertainty estimate computed from the approach described
    in Refs.\ \onlinecite{zuckerman-bias,bustamante-bias}.\\
    }
\end{table*}

Table \ref{tab-fkbp} shows the binding affinity results obtained
via Eq.\ \eqref{eq-dgbind}, with the PMF computed using both the
Hummer-Szabo approach of Eq.\ \eqref{eq-hz}
and the stiff-spring second cumulant approximation of Eq.\ \eqref{eq-gauss}.
The computational estimates of $\dgb$ for the target PMF
are in excellent agreement with experimental data.
The non-equilibrium estimates are typically within less than 4.0 kJ/mol 
of the target values.
Reference distances were chosen as $\refr=2.4$ nm
for both DMSO and BUQ, and
the value of the restraint free energy was found to be $\dga=-9.0$ kJ/mol
for FKBP-DMSO and $\dga=-7.8$ kJ/mol for FKBP-BUQ.
Finally the value of the apo integral in Eq.\ \eqref{eq-dgbind}
was computed analytically to be
10.6 kJ/mol for FKBP-DMSO and 10.7 kJ/mol for FKBP-BUQ.
The results show that non-equilibrium estimates for the FKBP-DMSO
system are more accurate than the FKBP-BUQ estimates
suggesting that the FKBP-BUQ system
requires more simulation time to converge.
Uncertainty estimates were obtained using both a bootstrap method
and the approach described in Refs.\
\onlinecite{zuckerman-bias,bustamante-bias}.

Table \ref{tab-fkbp} includes the standard deviation of the work values
$\sigma_W$ measured at the reference distance $\refr=2.4$ nm.
Previous studies have suggested that the optimal efficiency for
use of the Jarzynski relation is when the speed is slow enough
that $\sigma_W\approx 1.0\; k_BT \approx 2.5$ kJ/mol
\cite{hummer,crooks-pre,ytreberg-efficiency}.
Apparently the speeds attempted for the current study
were not slow enough to generate work values with such a small $\sigma_W$.
We note however, for the current study, the uncertainty
does not appear to correlate with $\sigma_W$.
Future studies will be carried out to determine if there
is an optimal pulling speed for these systems.

The results from Tab.\ \ref{tab-fkbp} suggest that use of
the Hummer-Szabo approach, while exact in the limit of infinite
sampling, is not feasible for the current study.
This is likely due to the fact that the pulling speeds used were too
fast to generated work values with $\sigma_W\approx 1.0\; k_BT$.
However, use of the approximate stiff-spring second cumulant expansion,
while approximate, does tend to improve the $\dgb$ estimates.
This is consistent with the recent study by Minh and McCammon
which determined that when similar speeds are utilized the
stiff-spring second cumulant expansion method performed better than
the other tested methods \cite{minh-jpcb-2008}.

We realize that the use of larger more flexible ligands may 
lead to difficulties in using the method suggested here.
This is due to the large number of possible conformations the
ligand may adopt in the apo state; all of which must be sampled
adequately to obtain an accurate PMF.
However, the method may be modified by including an additional
restraint to the RMSD of the ligand, thus restricting the conformational
freedom of the ligand.
The free energy of release from this RMSD restraint must then be included in
the binding affinity estimate \cite{roux-wham,roux-fkbp,mobley-restraint}.

\subsection{Note on simulation time}

Each non-equilibrium estimate in Table \ref{tab-fkbp} was generated
using a total simulation time of 216.0 ns
(1.0 ns position restrained + 4.0 ns equilibrium to generate
starting configurations + 200.0 ns for unbinding simulations
+ 11.0 ns for $\dga$ estimation).
Note however, that the unbinding simulations
were performed in parallel.
So, for example, at a speed of $2.0\times 10^{-4}$ nm/ps, twenty independent
10.0 ns simulations were performed in parallel.
Therefore, all the simulation data needed to compute $\dgb$ can
can be obtained in around a day of wall-clock time with the use
of a computer cluster.

\section{Conclusions}

We have demonstrated that non-equilibrium unbinding simulations
utilizing a physical pathway
can be used to generate estimates of the binding affinity
for the FKBP-DMSO and FKBP-BUQ systems studied here.
The non-equilibrium estimates are typically within less than 4.0 kJ/mol
of the target values; and the target values are within less than
1.0 kJ/mol of experiment.

Our results suggest that when the standard deviation of the work values
is larger than the optimal $\sigma_W\approx 1.0\; k_BT$
that the stiff-spring second cumulant expansion approximation provides
a better $\dgb$ estimate than the exact Hummer-Szabo method.

The importance of pursuing methods such as described here is
that such non-equilibrium approaches are trivially parallelizeable
since each unbinding simulation is performed independently.
Also, due to the use of a physical pathway, the method is simple
to implement in many existing simulation packages with little or
no modification to the software.

We note that the method described here is not expected to 
produce accurate binding affinities when the ligand is large and flexible.
In this case, it is necessary to extend the approach to include
additional restraints to the ligand during the unbinding
simulation to prevent large-scale fluctuations.
The contribution to the binding affinity from these additional
restraints must then be taken into account
\cite{roux-wham,roux-fkbp,mobley-restraint}.

The results obtained here suggest that non-equilibrium unbinding
simulations can be used to generate accurate estimates of
binding affinities.
Efficiency analysis and comparison to 
other methodologies will be carried out in future work.

\section*{Acknowledgments}
Funding for this research was provided by the University of
Idaho, Idaho NSF-EPSCoR and BANTech.
Computing resources were provided by IBEST at University of Idaho,
and by the TeraGrid Advanced Support Program.
FMY would like to thank Ronald White, David Mobley, and Daniel Zuckerman
for helpful discussion.

\bibliography{/home/marty/texmf/bibtex/fmy}

\end{document}